\documentclass[article]{aa}  

\usepackage{graphicx}
\usepackage{epstopdf}
\usepackage{txfonts}
\usepackage{ulem} 
\usepackage{amsmath,amsfonts,amssymb}
\usepackage{color}

\begin{document}

   \title{Population synthesis to constrain Galactic and Stellar Physics}

   \subtitle{I- Determining age and mass of thin-disc red-giant stars}

   \author{N. Lagarde
          \inst{1}, 
         A.C. Robin\inst{1}, C. Reyl\'e\inst{1}, and G. Nasello\inst{1}  
          }

   \institute{Institut UTINAM, CNRS UMR6213, Univ. Bourgogne Franche-Comt\'e, OSU THETA Franche-Comt\'e-Bourgogne, Observatoire de Besan\c con, BP 1615, 25010 Besan\c con Cedex, France \\
              \email{nadege.lagarde@utinam.cnrs.fr}
                }

   \date{Received 14 december 2016 / Accepted 5 February 2017}
 
  \abstract 
   {The cornerstone mission of the European Space Agency, Gaia, together with forthcoming complementary surveys (CoRoT, \textit{Kepler}, K2, APOGEE and Gaia-ESO), will revolutionize our understanding of the formation and history of our Galaxy, providing accurate stellar masses, radii, ages, distances, as well as chemical properties for a very large sample of stars across different Galactic stellar populations.}
      { Using an improved population synthesis approach and new stellar evolution models we attempt to evaluate the possibility of deriving ages and masses of clump stars from their chemical properties.}
   {A new version of the Besan\c con Galaxy model (BGM) is used in which new stellar evolutionary tracks are computed from the stellar evolution code STAREVOL. These provide global, chemical and seismic properties of stars from the pre-main sequence to the early-AGB. For the first time, the BGM can explore the effects of an extra-mixing occurring in red-giant stars. In particular we focus on the effects of thermohaline instability on chemical properties as well as on the determination of stellar ages and masses using the surface [C/N] abundance ratio.}
   {The impact of extra-mixing on $^3$He, carbon isotopic ratio, nitrogen, and [C/N] abundances along the giant branch is quantified. We underline the crucial contribution of asteroseismology to discriminate between evolutionary states of field giants belonging to the Galactic disc. The inclusion of thermohaline instability has a significant impact on $^{12}$C/$^{13}$C, $^3$He as well as on the [C/N] values. We clearly show the efficiency of thermohaline mixing at different metallicities and its influence on the determined stellar mass and age from the observed [C/N] ratio. We then propose simple relations to determine ages and masses from chemical abundances according to these models.} 
   {We emphasize the usefulness of population synthesis tools to test stellar models and transport processes inside stars. We show that transport processes occurring in red-giant stars should be taken into account in the determination of ages for future Galactic archaeology studies.} 
  

   \maketitle
%

\section{Introduction}

Galactic Archaeology explores  the formation and evolution of our Galaxy using the chemical properties, kinematics, and their dependency on age, of different stellar populations,  \citep{Freeman02,Turon08}. In this context, the determination of accurate stellar distances and ages  along the Galactic disc is crucial to improving our understanding of the Milky Way. \\

The Gaia space mission provided astrometry for more than 2 million stars and photometry for 1 billion stars, with typical uncertainties of about 0.3 mas for the positions and parallaxes, and about 1 mas/yr for the proper motions \citep[first data release,][]{GaiaDR1}.Asteroseismology data of red-giants stars observed by the space missions CoRoT \citep{BagFri06}, \textit{Kepler} \citep{Gilliland10} and K2 provide crucial constraints on the stellar properties such as masses, radii, and evolutionary states \citep[e.g.][]{Stello08,Mosser12a,Bedding11,Vrard16}, on the internal rotation profile \citep[e.g.][]{Mosser12b,Beck12}, as well as on the properties of helium ionization regions \citep{Miglio10}. Thanks to asteroseismology, masses of red-giant stars can be directly related to stellar interior physics and stellar evolution  \citep{Lebreton14a,Lebreton14b} allowing one to determine ages, without being limited to surface properties. 
Seismic data collected for more than 20,000 red-giant stars belonging to the Galactic-disc populations represent a huge sample to constrain stellar and Galactic physics \citep{Miglio13,Anders16}. A broad effort is ongoing with large spectroscopic surveys such as APOGEE \citep{APOGEE15, SDSSref}, ESO-Gaia \citep{Gilmore12}, RAVE \citep{RAVEDR1}, SEGUE \citep{SEGUE}, LAMOST \citep{LAMOST}, HERMES \citep{HERMES} and for the future with WEAVE, 4MOST and MOONS from which stellar parameters, radial velocities and detailed chemical abundances can be measured for CoRoT, \textit{Kepler}, and K2 targets.

    \begin{figure*}
   \centering
   \includegraphics[width=0.48\hsize]
   {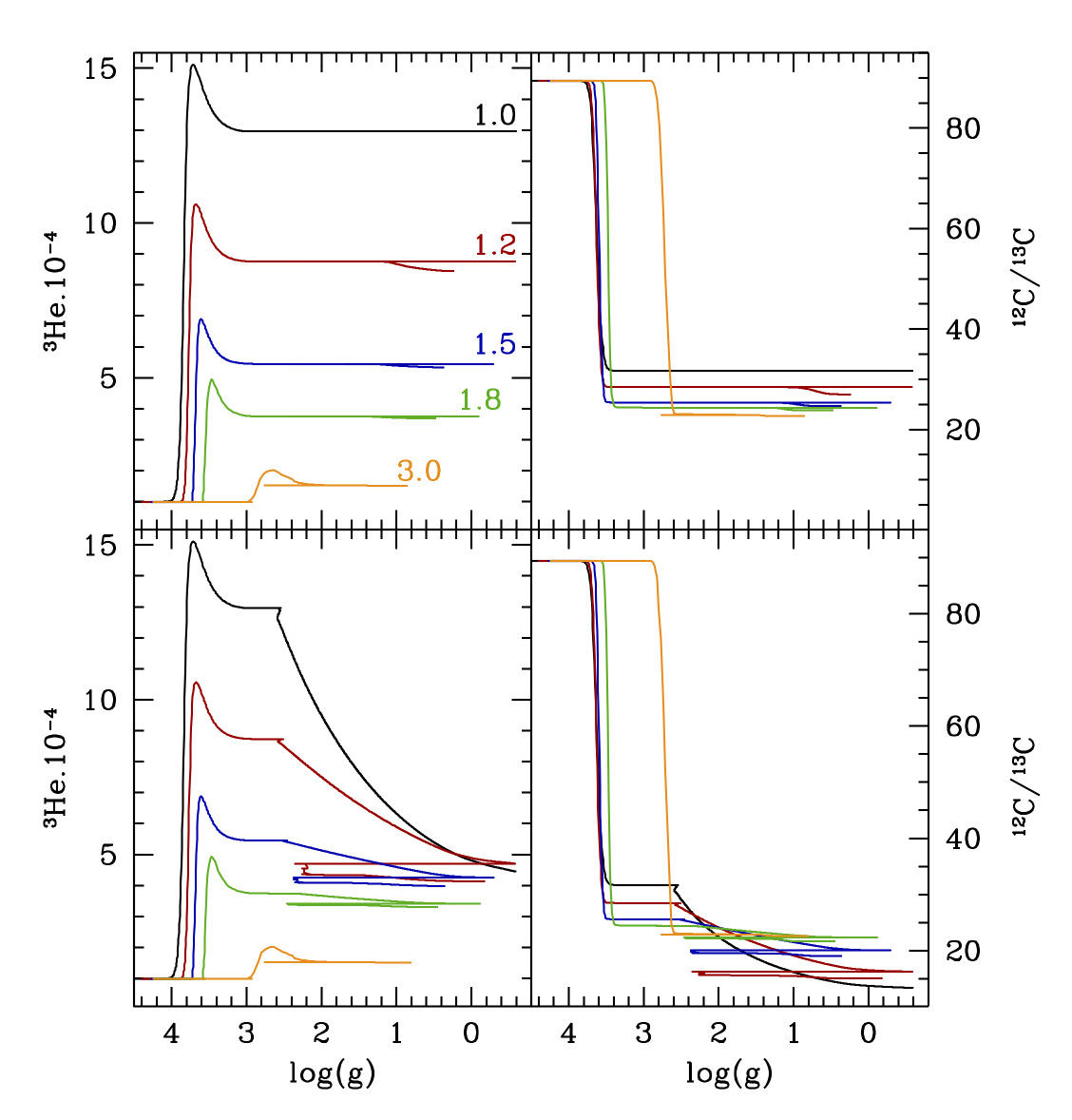}
   \includegraphics[width=0.48\hsize]
   {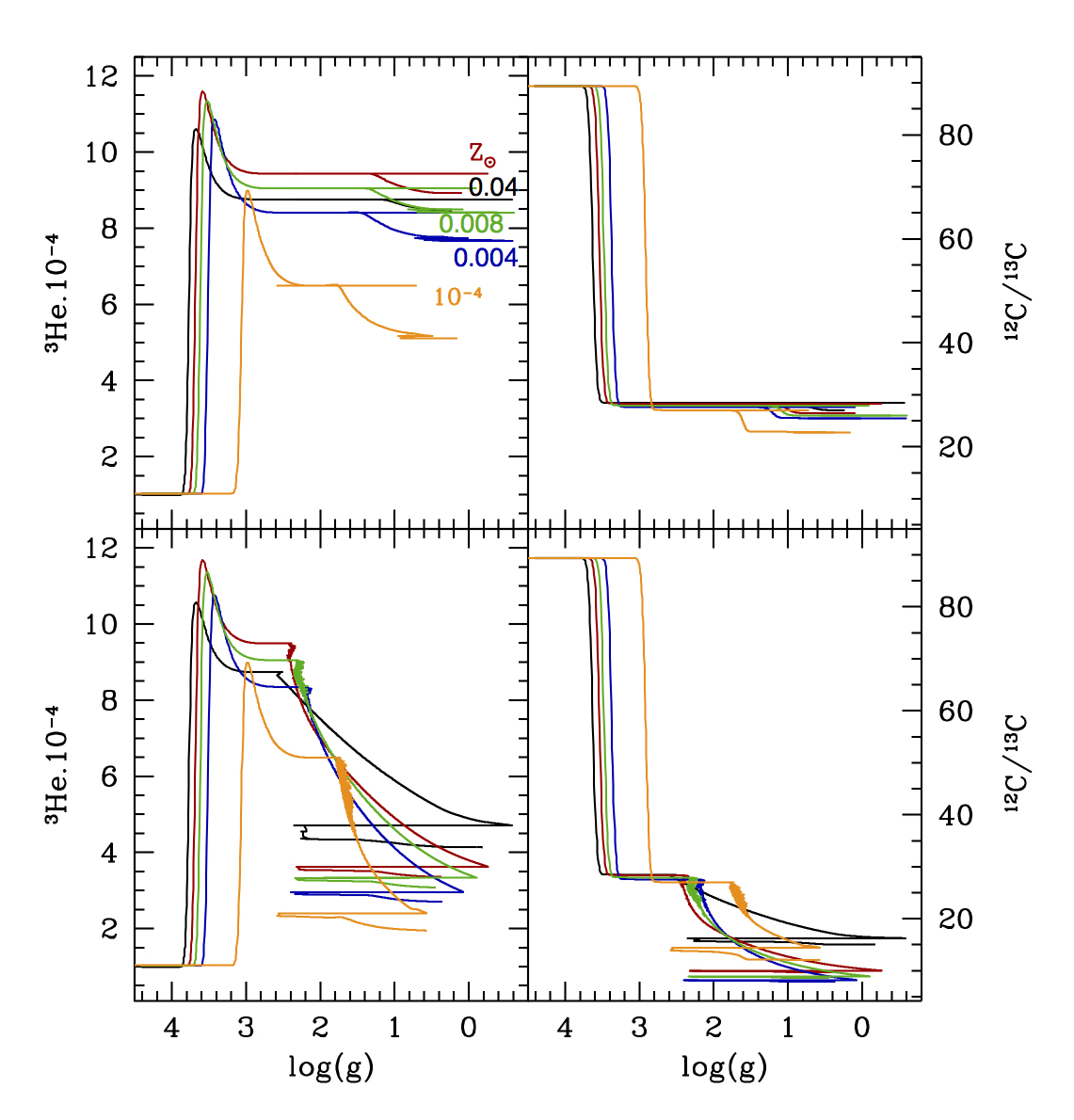}
      \caption{Theoretical evolution of $^3$He and carbon isotopic ratio at the stellar surface as a function of surface gravity for stellar models of : \textit{left panels} various masses (1.0 ; 1.2 ; 1.5 ; 1.8 ; 3.0 M$_{\odot}$ represented by a black, red, blue, green and orange solid line, respectively) at Z=0.04  ; \textit{right panels} various metallicities ([Fe/H]= 0.5 ; 0 ; -0.23 ; -0.54 ; -2.14 represented by a black, red, blue, green and orange solid line, respectively) at M=1.2 M$_\odot$. These models include the effects of thermohaline instability (bottom panels) and following standard evolution (top panels). These tracks are shown from the main sequence up to the early-AGB. }
         \label{Fig_Stars}
   \end{figure*}
    
To exploit their full potential,  it is crucial to perform a combined analysis of these different kinds of observations. The population synthesis approach is a powerful tool for such analysis, allowing the computation of mock catalogues under various model hypothesis, and to statistically compare them with any type of large survey data. The Besan\c con Galaxy model (hereafter BGM) is a stellar population synthesis model \citep{Robin03, Czekaj14} intended to meld the formation and evolution scenarii of the Galaxy, stellar formation and evolution theory, models of stellar atmospheres, as well as dynamical constraints, in order to make a consistent picture of the Galaxy in comparison with available observations (photometry, asteroseismology, astrometry, and spectroscopy) at different wavelengths. 

To benefit from the combination of recent asteroseismic and spectroscopic surveys, we updated the evolutionary tracks that are used as inputs in the BGM. These stellar evolution models are computed with the code STAREVOL \citep[e.g.][]{Lagarde12a}, which follows the global, chemical and seismic properties of stars all along their evolution. This is done from the pre-main sequence (along the Hayashi track) to the early-asymptotic giant branch (early-AGB). 
These models include the effects of different transport processes such as thermohaline instability (discussed in this paper). \\

The aim of this series of papers is to investigate the impacts of different hydrodynamic processes that occur inside the stars on the chemical properties of Galactic stellar populations. In the context of the interpretation of large spectroscopic surveys such as APOGEE, or Gaia-ESO, we will perform comparison between theoretical synthetic populations and spectroscopic surveys in forthcoming papers (Lagarde et al. in prep. Part II). In this paper, we focus on presenting and discussing the implementation in the Besan\c con Galaxy model of stellar evolution models that include the effects of thermohaline instability on global, chemical and seismic properties. We discuss the determination of stellar ages and masses from the [C/N] ratio. We plan to focus on the effects of rotation and the implementation of different prescriptions for thermohaline instability on stellar ages and chemical properties in a separate forthcoming paper (Part III). 
In Sect.2  we present the input physics of our stellar evolution models and briefly recall the impacts of thermohaline instability on the chemical properties of giant stars. The Besan\c con Galaxy Model is presented in Sect.3, while in Sect. 4 we present synthetic stellar populations, taking into account the effects of thermohaline mixing or following the  standard prescription. We also present the new quantities that can be simulated by the BGM. In Sect. 5, we discuss the determination of stellar ages and masses from the surface carbon and nitrogen abundances, and we provide relations for three metallicity ranges. We conclude and explore some perspectives in Sect. 6.

\section{Stellar evolution models}

\subsection{Description of stellar evolution models}

\begin{table}
\caption{Stellar Evolution models}             
\label{stellarevo}      
\centering                          
\begin{tabular}{c c c}        
\hline\hline                 
Metallicity sets   & Stellar mass &  Mixing \\ 
Z ([Fe/H], [$\alpha$/Fe]) &(M$_\odot$)  &\\   
\hline                        
   Z=0.04 (0.51, 0)  &   &     \\      
   Z=0.0134 (0, 0) & 0.6, 0.7, 0.8, 0.9, 1.0, & Standard,  \\
   Z=0.008 (-0.23, 0) & 1.1, 1.2, 1.3, 1.4, 1.5, & Thermohaline \\
   Z=0.004 (-0.54, 0) & 1.6, 1.7, 1.8, 1.9, 2.0, & \\
   Z=0.0001 (-2.15, 0.3) &  & \\ 
\hline                                   
\end{tabular}
\end{table}

Stellar evolution models are computed with the code STAREVOL \citep[e.g.][]{Lagarde12b} for a range of masses between 0.6~M$_\odot$ and 6.0~M$_\odot$ at five metallicities Z=0.04 ([Fe/H]=0.51), Z=0.0134 ([Fe/H]=0), Z=0.008 ([Fe/H]=-0.23), Z=0.004 ([Fe/H]=-0.54) and Z=0.0001 ([Fe/H]=-2.14). These models are computed from the pre-main sequence to the early-AGB phase. We use the same main physical ingredients that are used and fully described in \citet{Lagarde12b}, except for:

\begin{itemize}
\item the solar mixture, that comes from \citet{Asplund09};

\item the treatment of convection, that is based on a classical mixing length formalism with $\alpha_{MLT}=$1.6264 recovered from solar-calibrated models that include neither atomic diffusion nor rotation.

\end{itemize}

As discussed in \citet{Lagarde12a}, these stellar evolution models follow the global asteroseismic properties using the scaling relations and asymptotic relations \citep{Tassoul80}, i.e. the large frequency separation, the frequency corresponding to the maximum oscillation power, the asymptotic period spacing of g-modes, and different acoustic radii.

To quantify the effects of transport processes on the chemical properties of stellar populations in our Galaxy, we computed stellar evolution models assuming: (1) standard models (no mixing mechanism other than convection), (2) models that include the effects of thermohaline instability induced by $^{3}$He-burning. This instability develops as long thin fingers whose aspect ratio is consistent with prediction by \citet{Ulrich72} and with the laboratory experiments \citep{Krish03}. Although the efficiency of thermohaline instability in stellar interiors is still discussed in the literature by hydrodynamical simulations \citep[][ Prat et al in prep.]{Denissenkov10, Traxleretal11}, we adopt an aspect ratio equal to 5, which might correspond to the maximum efficiency of this instability. \\

As discussed by \citet{ChaLag10} and \citet{Angelou11,Angelou12}, this mechanism has a crucial effect on surface chemical properties of red-giant branch stars. Its effect is consistent with most of spectroscopic observations of low-mass and intermediate-mass stars, especially in reproducing the low-$^{12}$C/$^{13}$C and the [C/N] ratio in open clusters \citep[e.g.][]{Tautvaisiene16,Drazdauskas16, Tautvaisiene15,Tautvaisiene13,Mikolaitis10,Smiljanic09} as well as in CoRoT targets \citep{Morel14,Lagarde15}.
It has also a very significant impact on the chemical evolution of light elements in the Milky Way \citep{Lagarde11} allowing the resolution of the long-standing ''$^3$He-problem" in our Galaxy \citep{Lagarde12b}. \\

In the following section, we briefly describe the evolution of stars and present the effects of thermohaline mixing on the surface abundances of $^3$He and carbon isotopic ratio drawing the stellar evolution.

    \begin{figure}
   \centering
   \includegraphics[width=\hsize,clip=true,trim= 0.8cm 0.5cm 0.8cm 0.8cm]{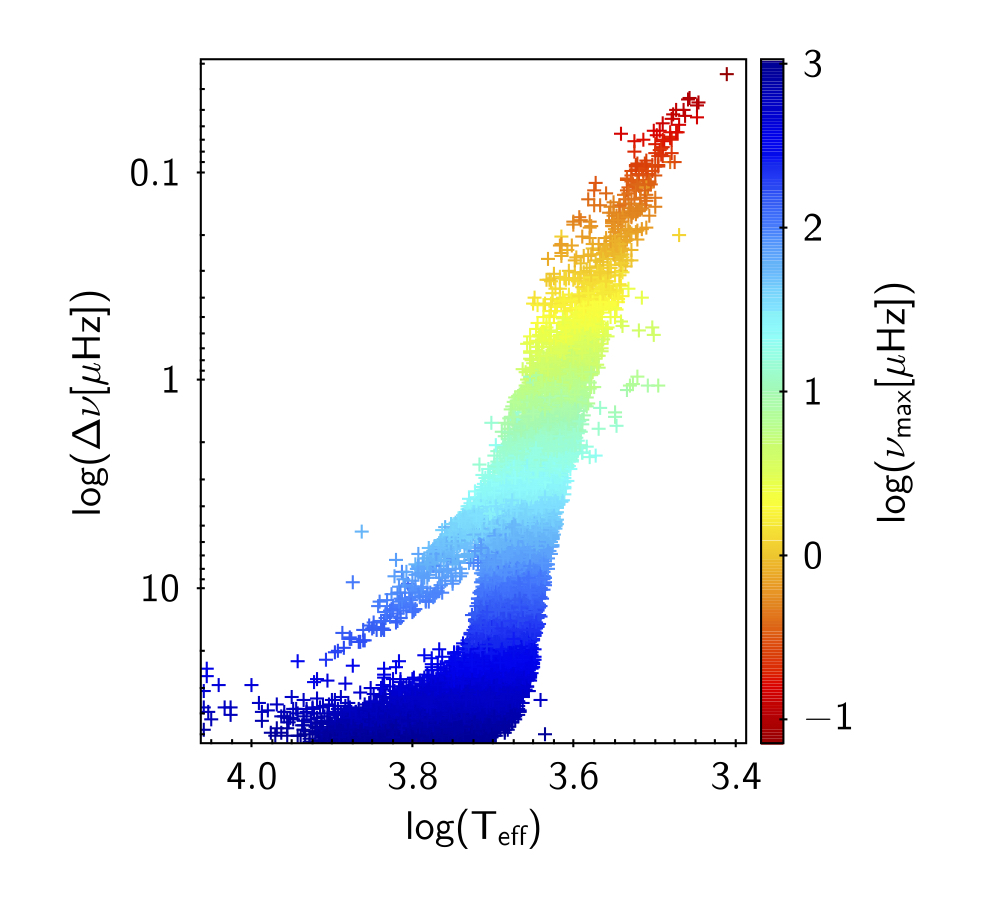}
      \caption{The large separation, $\Delta \nu$, as a function of the effective temperature, T$_{\rm{eff}}$, for synthetic population computed with the BGM. Colours indicate the $\nu_{\rm{max}}$, the frequency at which the power spectrum is maximum.}
         \label{Fig_sismo}
   \end{figure}

\subsection{Chemical evolution of low- and intermediate-mass giant stars}
Figure \ref{Fig_Stars} presents the theoretical evolution of Helium-3 and carbon isotopic ratio for different initial masses and metallicities from the main sequence to the early-AGB.
 
\subsubsection{Standard evolution}

After the main sequence, a star experiences core contraction to increase its central temperature, then evolves rapidly toward the red giant branch (RGB). During this phase, called the ''Sub giant branch`` (SGB) when the stars cross HR diagram almost horizontally, the convective envelope grows in mass and deepens inside the star. This is the first dredge-up, where the convective envelope is diluted with hydrogen-processed material, inducing changes of the surface abundances. This episode occurs at higher luminosity when the metallicity decreases as well as when the stellar mass increases. The surface mass fractions of $^7$Li, $^9$Be, $^{12}$C and $^{18}$O decrease while those of $^{3}$He, $^4$He, $^{13}$C, $^{14}$N, and $^{17}$O increase, implying a decrease at the surface of the isotopic ratios $^{12}$C/$^{13}$C, and $^{12}$C/$^{14}$N. Figure \ref{Fig_Stars} clearly shows the signature of this episode (e.g. at log(g)$\sim$3.8 for 1.2M$_{\odot}$).

As shown by \citet{Charbonnel94}, the first dredge-up efficiency (in terms of maximum depth of the convective envelope) decreases with decreasing metallicity. For a given stellar mass, Fig.\ref{Fig_Stars} (right panel) shows that the depletion of $^{12}$C/$^{13}$C and the increase of $^3$He begin at lower gravity when the metallicity decreases. The chemical variations during the first dredge-up depend on the initial stellar mass as well as on the metallicity.
For a given metallicity, when the initial stellar mass increases, Fig.\ref{Fig_Stars} (left panels) shows that the post dredge-up values of carbon isotopic ratio and $^3$He decrease, as well as at the given stellar mass when the metallicity decreases (right panel). This dependence is true for $^{12}$C/$^{14}$N, as well as for oxygen isotopic ratios. This is due to the convective envelope reaching deeper regions during the first dredge-up. 
From the sub giant branch to the RGB tip, the temperature at the base of the convective envelope is always insufficient to activate nuclear reactions. As a consequence, in standard
models (top panels of Fig. \ref{Fig_Stars}) the only change in surface abundances is due to the first dredge-up on the SGB; after this episode the surface abundances remain constant along the RGB once the convective envelope recedes and until the second dredge-up occurs at the end of the helium burning phase.

\begin{figure*}
   \centering
   \includegraphics[width=0.4\hsize,clip=true,trim= 0.8cm 0.5cm 0.8cm 0.8cm]{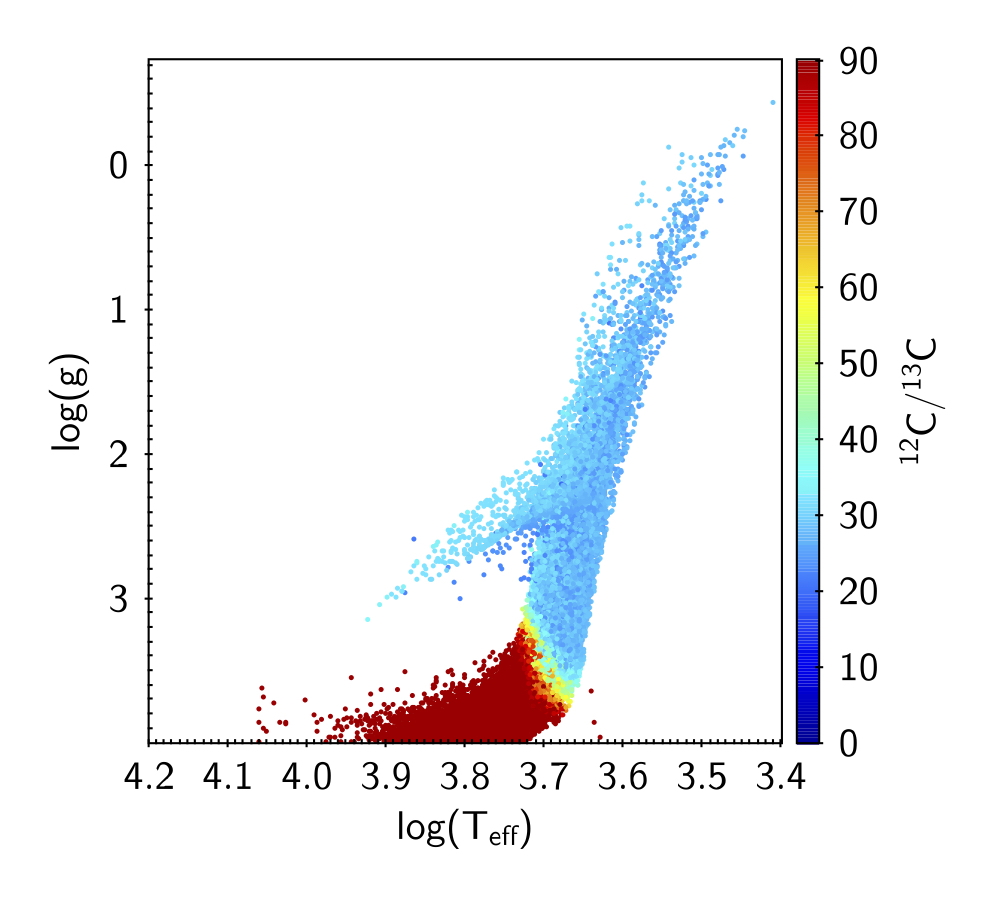}
   \includegraphics[width=0.4\hsize,clip=true,trim= 0.8cm 0.5cm 0.8cm 0.8cm]{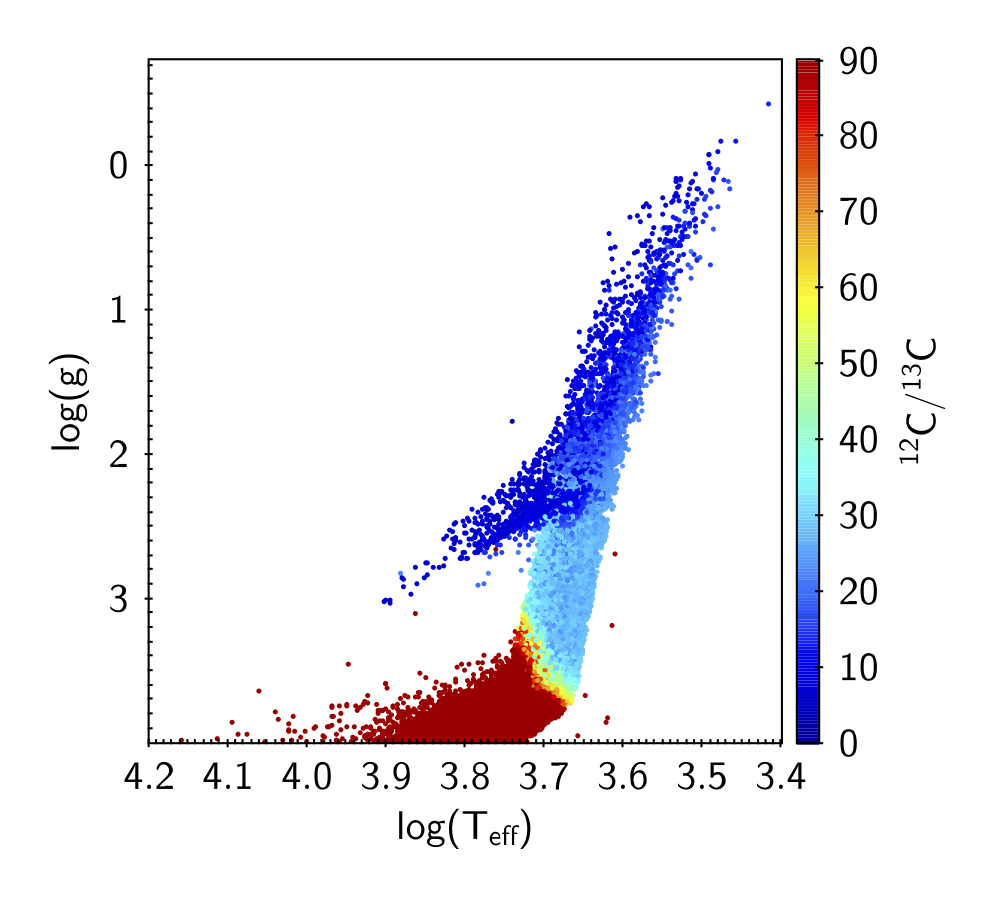}
      \caption{Surface gravity as a function of effective temperature for synthetic populations computed with the BGM including the effects of thermohaline instability (right panel) or not (left panel). The colour code represents the carbon isotopic ratio at the surface of stars in the thin disc.} 
         \label{Fig_popc1213}
   \end{figure*}

\subsubsection{Thermohaline instability during the RGB}
 
Thermohaline mixing has recently been flagged as the main mechanism that can govern the
photospheric chemical composition of low-mass bright giant stars \citep{ChaZah07a, ChaLag10}. In such stars, thermohaline instability is a double-diffusive instability induced by the molecular weight inversion created by the $^3$He($^3$He,2p)$^4$He reaction in the external region of the hydrogen-burning shell \citep{Eggleton06,Eggleton08}. 
As discussed by \citet{ChaZah07a} and \citet{ChaLag10}, this instability is expected to set in after the first dredge-up when the star reaches the RGB-bump (at log(g)$\sim$2.5 on bottom panels of Fig.\ref{Fig_Stars}). In terms of stellar structure, the RGB bump corresponds to the moment when the hydrogen-burning shell encounters the chemical discontinuity created by the convective envelope at its maximum penetration during the first dredge-up. When the hydrogen burning shell (which provides the stellar luminosity on the RGB) reaches the H-rich previously mixed zone, the corresponding decrease in molecular weight of the H-burning layers induces a drop in the total stellar luminosity, creating a bump in the luminosity function \citep[i.e. ][]{FusiPecci90, Charbonnel94, Charbonnel98}.

Thermohaline mixing induces a decrease of $^3$He at the stellar surface after the RGB-bump. $^{13}$C and $^{14}$N diffuse outwards, while $^{12}$C diffuses inwards \citep[see for more details][]{ChaLag10}, implying a decrease of $^{12}$C/$^{13}$C and [C/N]. The efficiency of thermohaline instability decreases when the stellar mass or stellar metallicity increase (see Fig. \ref{Fig_Stars}). In addition, thermohaline mixing induces a slight decrease of $^3$He \citep{Lagarde11} and increase of nitrogen at lower metallicity during the central helium burning before the second dredge-up episode. \\

Thermohaline instability does not change the evolutionary tracks in the HR-diagram and has not impact on stellar ages because it has not significant effect on the stellar structure.

 \begin{figure*}
   \centering
\includegraphics[width=8cm,trim= 0.5cm 0.5cm 0.8cm 0.4cm]{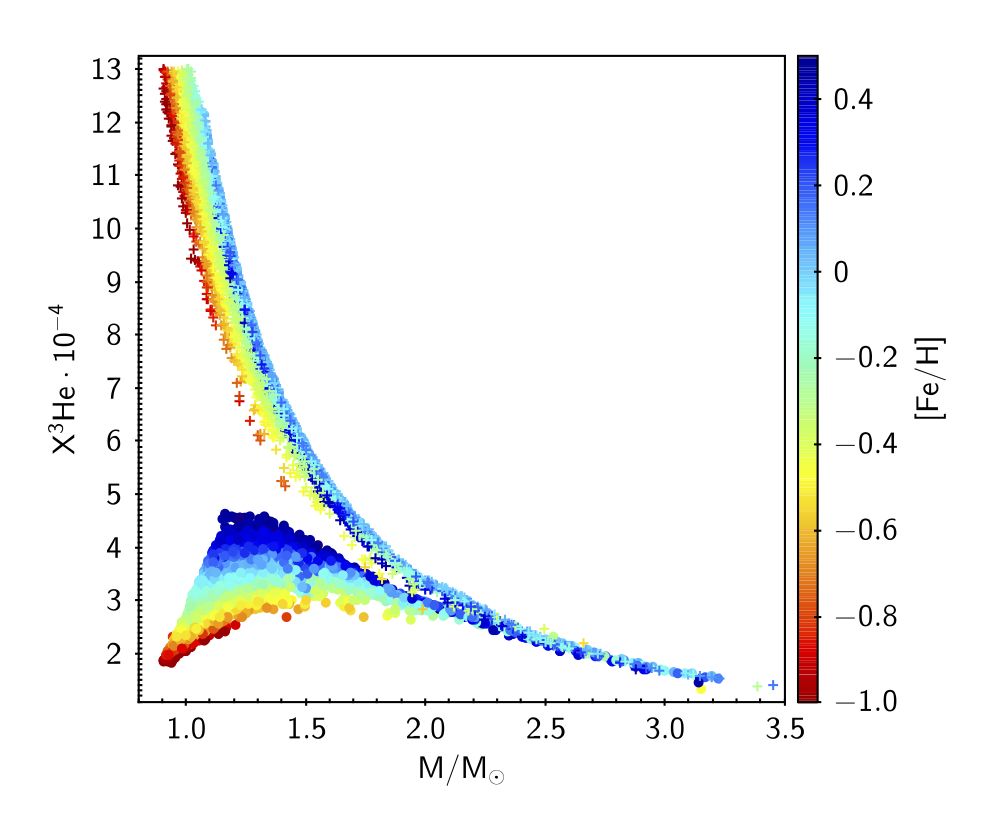}
\includegraphics[width=8cm,trim= 0.5cm 0.5cm 0.8cm 0.4cm]{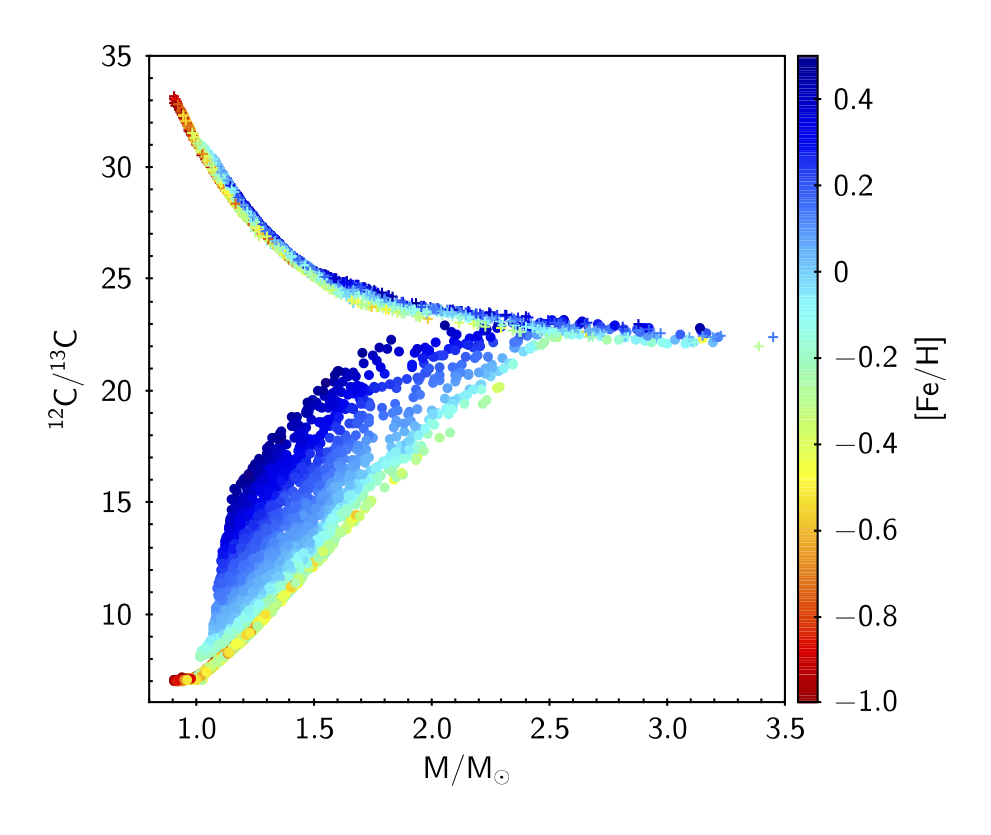}\\
\includegraphics[width=8cm,trim= 0.5cm 0.5cm 0.8cm 0.4cm]{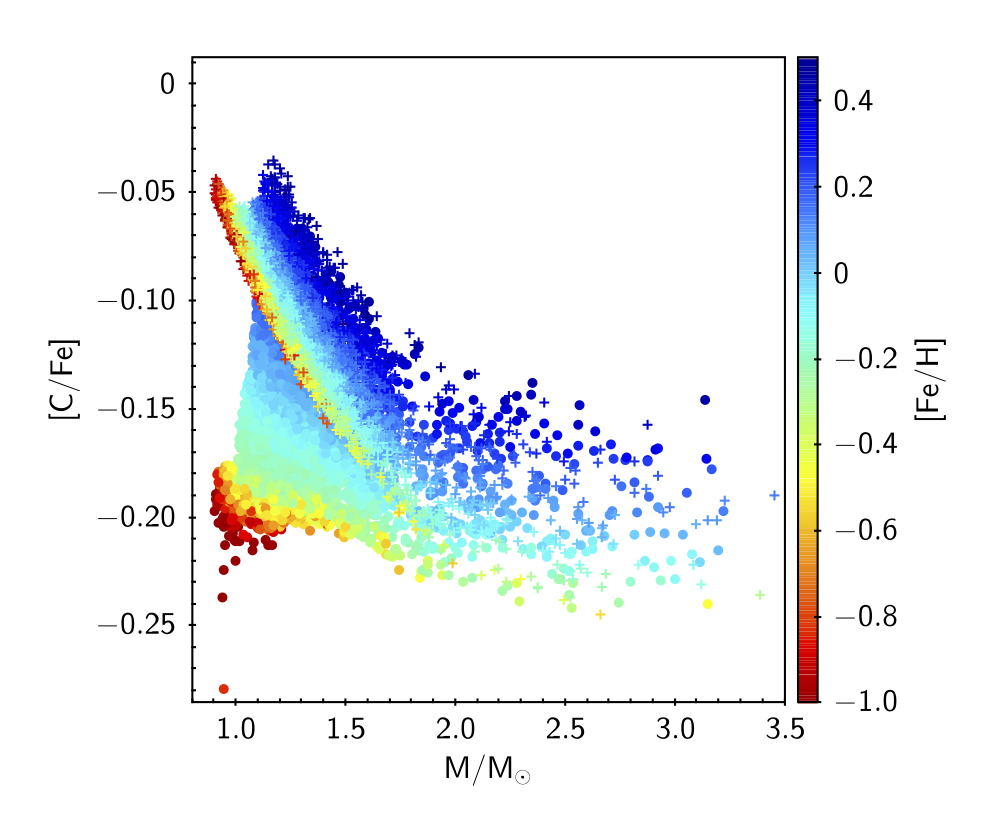}
\includegraphics[width=8cm,trim= 0.5cm 0.5cm 0.8cm 0.4cm]{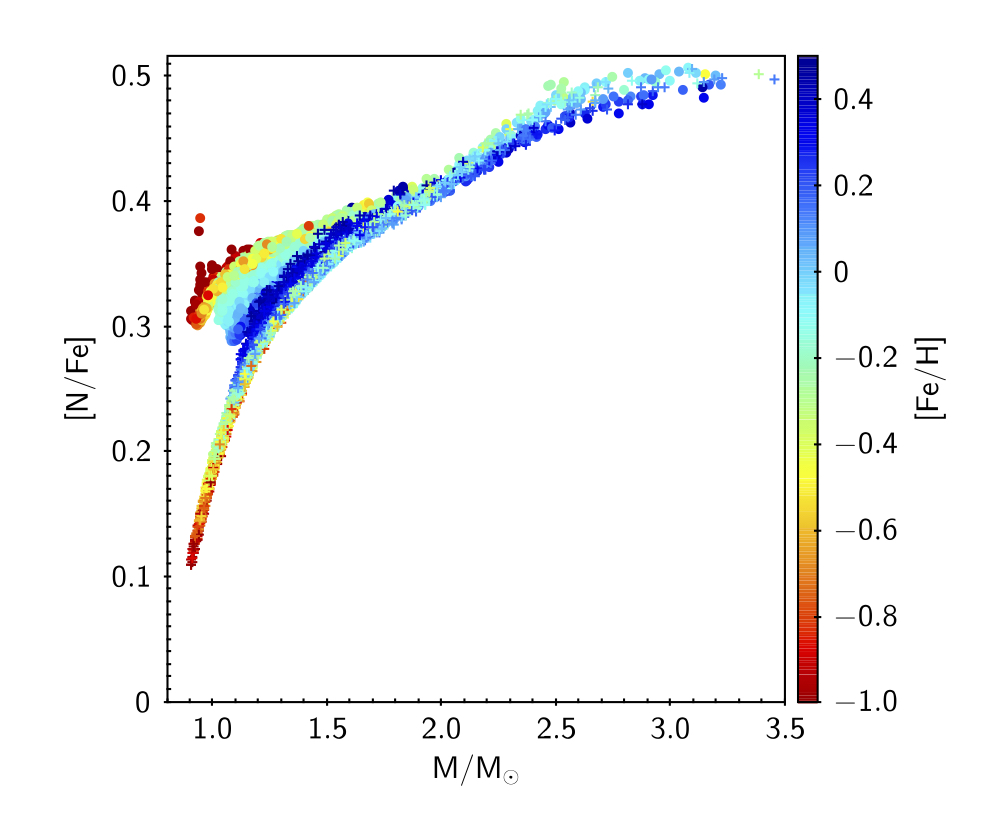}
         \caption{Surface abundance of $^3$He (in mass fraction, top-left panel), $^{12}$C/$^{13}$C  (top-right panel), [C/Fe] (bottom-left panel), and [N/Fe] (bottom-right panel) as a function of stellar mass for synthetic thin disc computed with the BGM. Stars have been selected to be in the clump according to their $\Delta\Pi_{\ell=1}$. The colour code represents the metallicity of stars. Simulations including or not the effects of thermohaline instability are represented by colour-dots and crosses respectively.}  
         \label{Fig_popClump}
   \end{figure*}

\section{The Besan\c con Galaxy Model}
The Besan\c con Galaxy Model (hereafter BGM) is a model using the population synthesis approach that simulates observations of the sky with errors and biases. It is based on assumptions and a scenario for the Galaxy formation and evolution that reflect the present knowledge about the Milky Way. 
Four stellar populations are considered: a thin disc, a thick disc, a bar, and a halo, with each stellar population having a specific density distribution. The stellar content of each population is modeled through an Initial Mass Function (IMF) and a Star Formation History (SFH), which can differ from one population to the other. To compute the stellar distribution at a given time, the stars generated by these IMFs and SFHs follow evolutionary tracks. When they have reached their final age (the present time) their astrophysical parameters (mass, age, T$_{\rm{eff}}$, log(g), metallicity, abundances...) are stored and used to compute their observational properties, using atmosphere models, and assuming a 3D extinction map describing the interstellar extinction they suffer. A dynamical model is used to compute radial velocities and proper motions. In the simulations presented here, we make use of the 3D extinction map from \cite{Marshall2006}.

For the time being, the metallicity of each star is the initial metallicity of the gas when it was born, which is estimated assuming a time dependent metallicity from \cite{Haywood2008}, and a radial metallicity gradient of -0.7 dex/kpc. In the future a proper chemical evolution model will be incorporated.

The model takes into account the stellar binarity, by generating secondary components at their birth, assuming a binarity probability depending on mass (see \cite{Czekaj14} for more details). Then, according to the estimated spatial resolution of the observations to simulate, stars in systems are merged or not (that is, their fluxes are added if the stars are too close, projected on the plane of the sky).

In the present version of the model, only the thin disc is modelled this way, while the thick disc, bar and stellar halo are generated using a Hess diagram computed from a fixed evolutionary scheme as described in \cite{Robin2003}, which does not yet take into account binarity.

The stellar densities in different regions of the Galaxy are modulated by density laws for each population, which are given in detail in \cite{Robin2003} for the thin disc, \cite{Robin2012a} for the bar, and \cite{Robin2014} for the thick disc and halo.

In the model version presented here, we use the evolutionary tracks described in Sect.2, instead of the Padova tracks used in \cite{Czekaj14}, from which surface abundances and asterosismic parameters are computed for each simulated star. For stars of mass lower than 0.6 solar mass, \cite{Chabrier1997A&A...327.1039C} evolutionary tracks are still used, although those new properties are not computed yet.

\section{Simulations of synthetic populations}
As discussed in Part 2 and in the literature \citep{Palacios06,Lagarde12a,Bossini15}, transport processes occurring in stellar interiors have significant impact on global (e.g. luminosity, effective temperature, age), chemical, and seismic properties. Population syntheses are powerful tools to study these processes using survey data. In the context of our study, we discuss the impact of thermohaline mixing on the properties of thin disc giants observed by asteroseismic and spectroscopic surveys. We shall consider in a future paper the thick disc population, which differs from the thin disc by its [$\alpha$/Fe] abundance ratio. For the present study we only consider stellar models according to solar $\alpha$-abundance (i.e. [$\alpha$/Fe]=0).
In order to estimate the impact of these stellar models on the populations observed, we performed simulations with the BGM using these new stellar evolutionary tracks and assuming the IMF and SFR given in \cite{Czekaj14}. Simulations are performed, for example, in the \textit{Kepler} field. They provide astrophysical parameters, asteroseismic properties and surface abundances for 54 stable and unstable species. We study in this section the impact on the properties of the simulated fields of this new models.
   
\subsection{Asteroseismic properties}

The detection of solar-like oscillations  with the space missions CoRoT and \textit{Kepler} provides powerful constraints on stellar mass and radius of giants stars. With the asymptotic period spacing of gravity modes, these observations provide information on stellar structure \citep{Mosser12a, Lagarde12a}, as well as constraints on transport processes \citep{Lagarde16}. Using the scaling relations \citep{Tassoul80, Mosser10, Belkacem11}, the asteroseismic parameters of large separation, $\Delta\nu$, and frequency of maximum oscillation power, $\nu_{\rm{max}}$ are directly related to stellar radii and masses:
\begin{equation}
\frac{M}{M_\odot} \approx  \left ( \frac{\nu_{max}}{\nu_{max,\odot}}\right )^{3} \left ( \frac{\Delta \nu}{\Delta \nu_{\odot}}\right )^{-4} \left ( \frac{T_{\rm{eff}}}{T_{\rm{eff},\odot}}\right )^{3/2} 
\end{equation}
\begin{equation}
\frac{R}{R_\odot} \approx  \left ( \frac{\nu_{max}}{\nu_{max,\odot}}\right )  \left ( \frac{\Delta \nu}{\Delta \nu_{\odot}}\right )^{-2} \left ( \frac{T_{\rm{eff}}}{T_{\rm{eff},\odot}}\right )^{1/2}  
\end{equation}

Solar reference values are $\Delta\nu_{\odot}$=135.1$\mu$Hz; $\nu_{\rm{max},\odot}$=3090$\mu$Hz, and T$_{\rm{eff},\odot}$=5777 K. \citet{Miglio12} and \citet{Mosser13a} proposed a relative correction to the scaling relations between red clump and RGB stars. This should affect the mass and radius determinations of clump stars up to $\sim$10\% and $\sim$6\% respectively. Different temperature scales have been tested in \citet{Miglio12} and agree within the quoted uncertainty range.

The population synthesis provides the seismic properties such as the large separation, the frequency with the maximum amplitude, or the asymptotic period spacing of g-modes for stars in the thin disc (see Fig.\ref{Fig_sismo}). 
 It is well known that the age of giant stars can be approximated by their lifetime on the main sequence, which depends on the stellar mass and metallicity. In addition, asteroseismology provides accurate stellar radii and thus allows the determination of distances. These observations of giants belonging to the different populations of the Milky Way provide additional constraints to study the formation and evolution of our Galaxy \citep{Miglio13}. To fully exploit the potential of these observations, it is crucial to combine them with spectroscopic surveys allowing the following of the chemical properties of the stars within the different stellar populations.

\subsection{Surface chemical properties}

Figure \ref{Fig_popc1213} shows the evolution of carbon isotopic ratio along the log(g) vs T$_{\rm{eff}}$ diagram in a simulation computed with the BGM. This figure clearly shows the impact of thermohaline instability on the $^{12}$C/$^{13}$C. The brighter-RGB and clump stars have a lower $^{12}$C/$^{13}$C at the surface when thermohaline instability is included.
   
    \begin{figure*}
   \centering
   \includegraphics[width=0.48\hsize,clip=true,trim= 0.8cm 0.5cm 0.8cm 0.4cm]{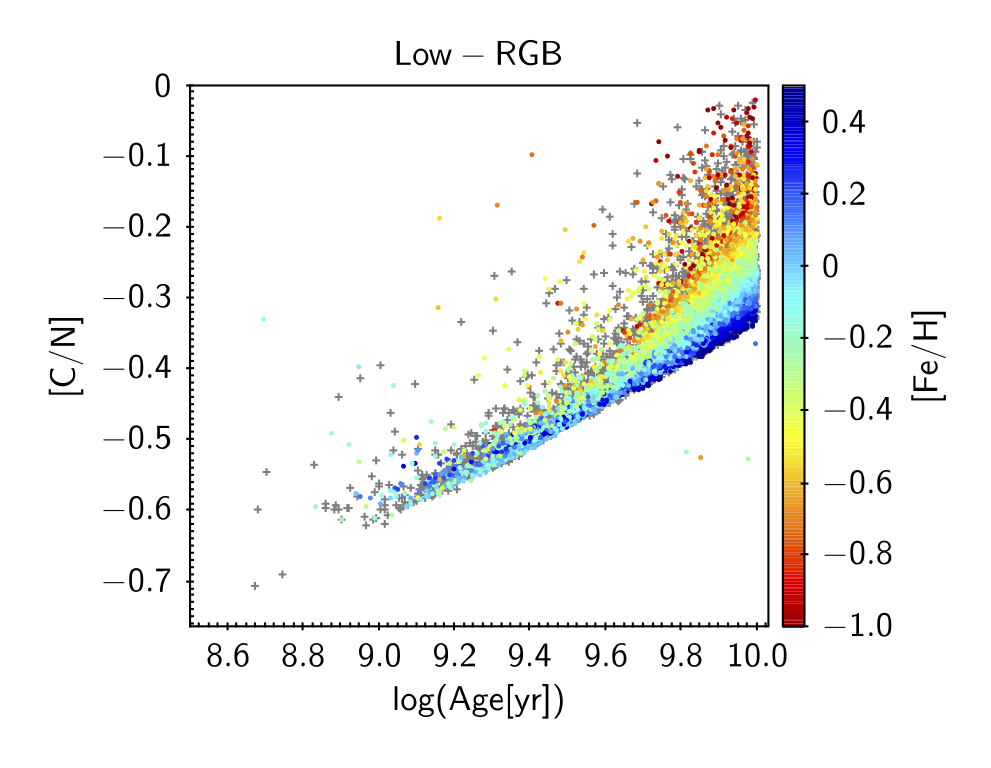}
   \includegraphics[width=0.48\hsize,clip=true,trim= 0.5cm 0.5cm 0.8cm 0.4cm]{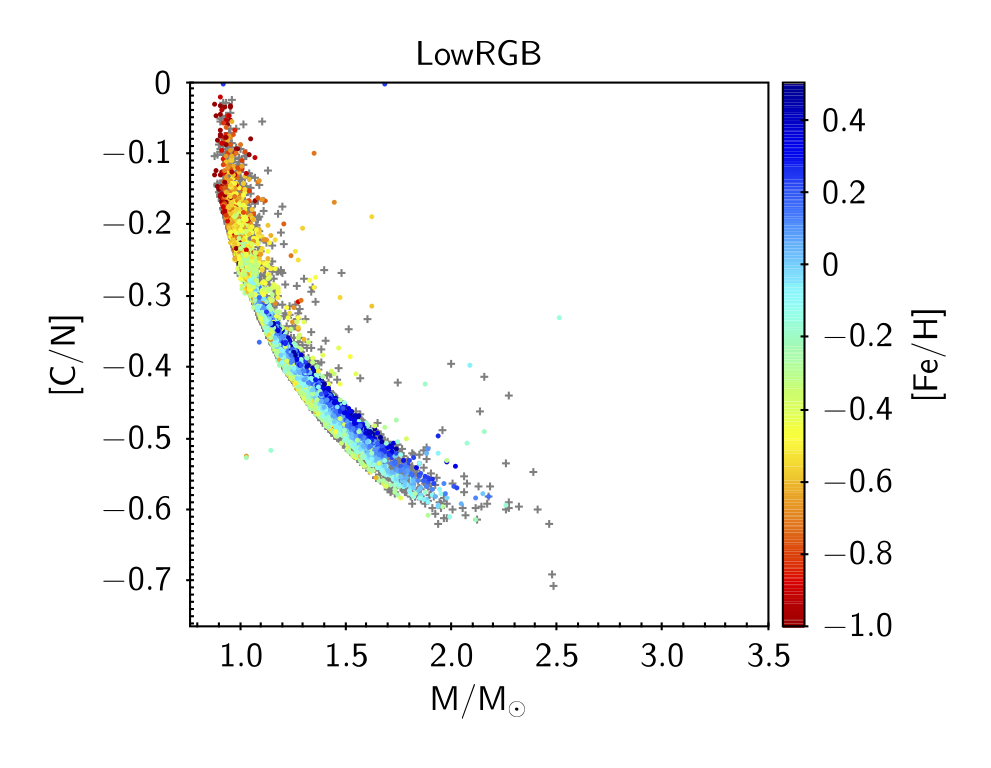}
   \includegraphics[width=0.48\hsize,clip=true,trim= 0.8cm 0.5cm 0.8cm 0.4cm]{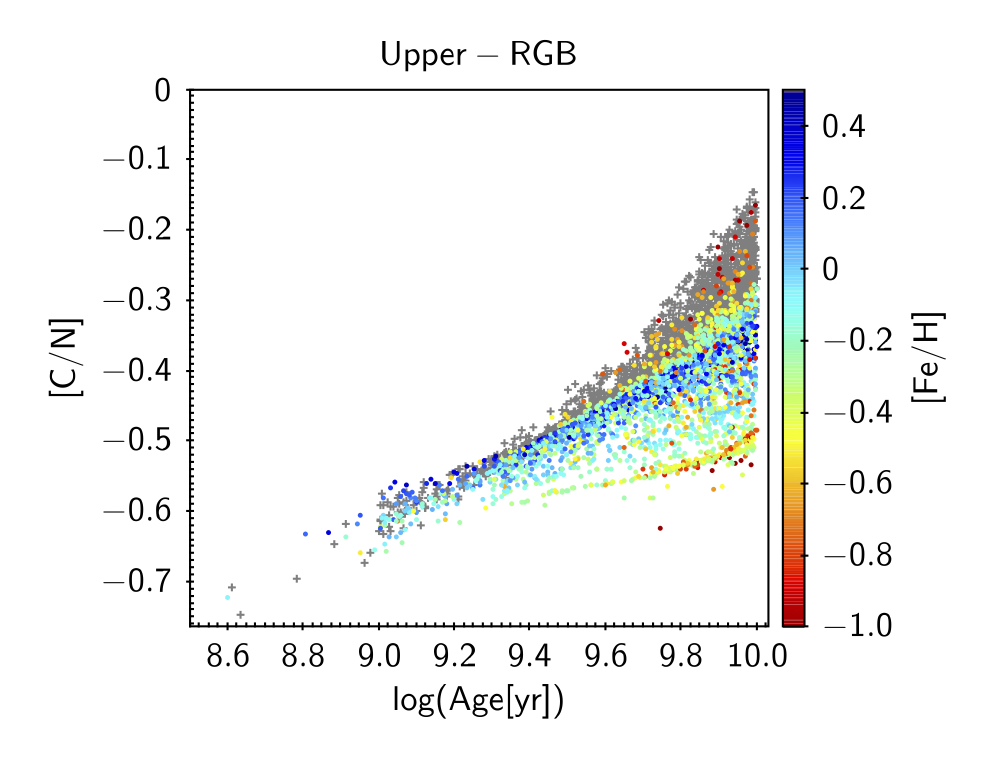}
    \includegraphics[width=0.48\hsize,clip=true,trim= 0.5cm 0.5cm 0.8cm 0.4cm]{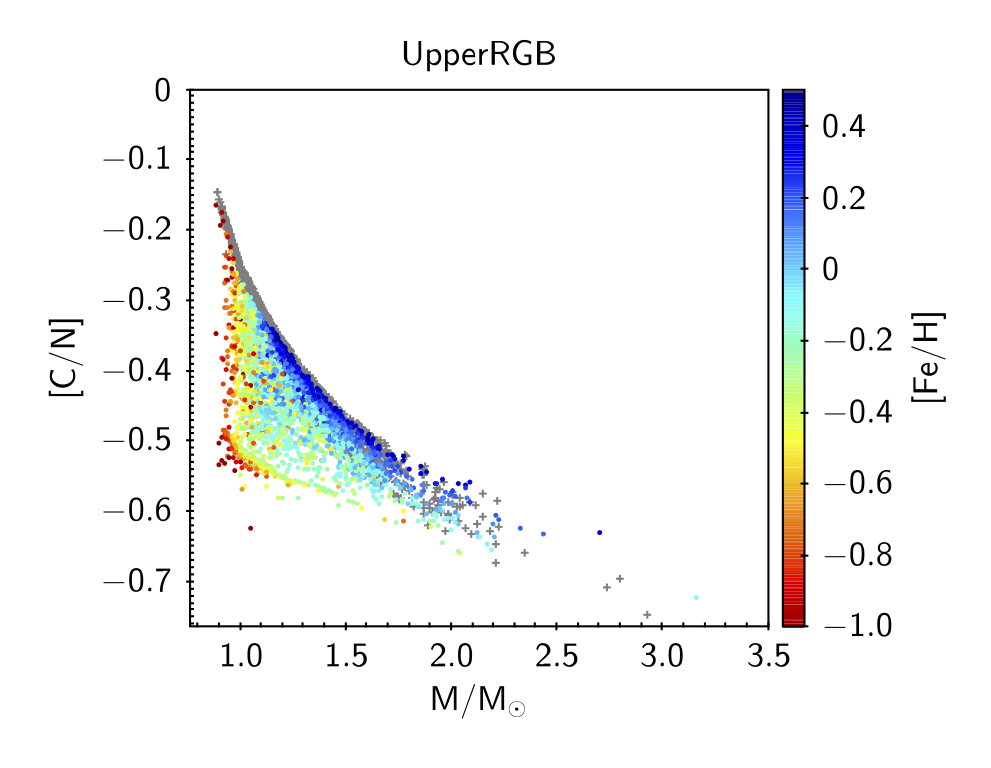}
   \includegraphics[width=0.48\hsize,clip=true,trim= 0.8cm 0.5cm 0.8cm 0.4cm]{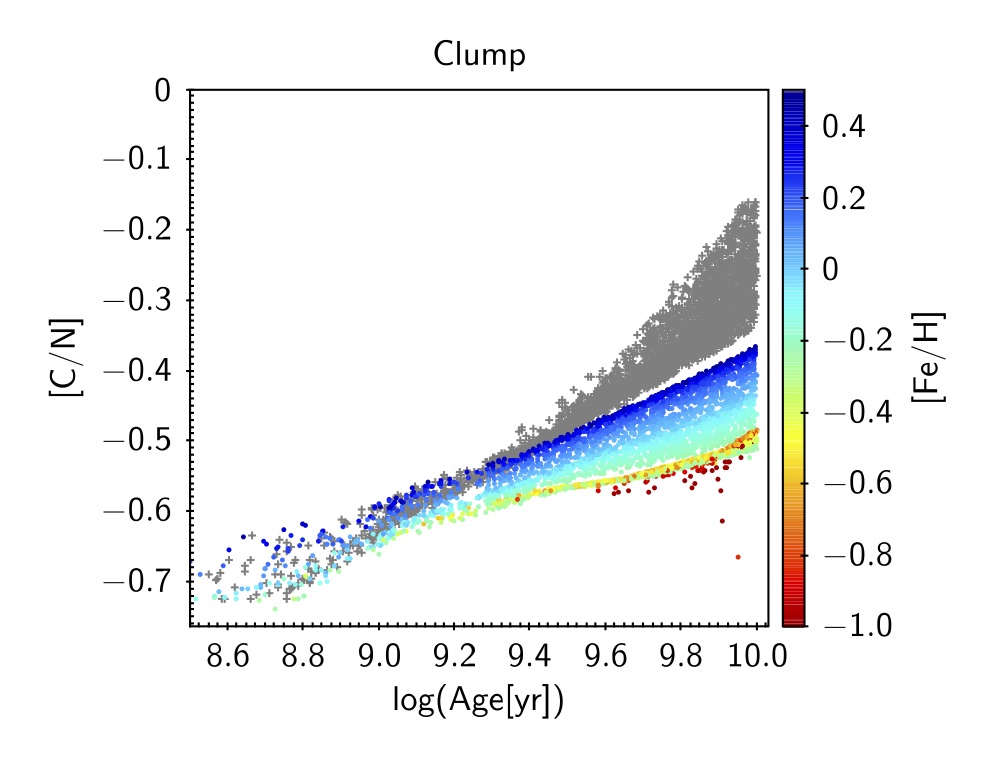} 
   \includegraphics[width=0.48\hsize,clip=true,trim= 0.5cm 0.5cm 0.8cm 0.4cm]{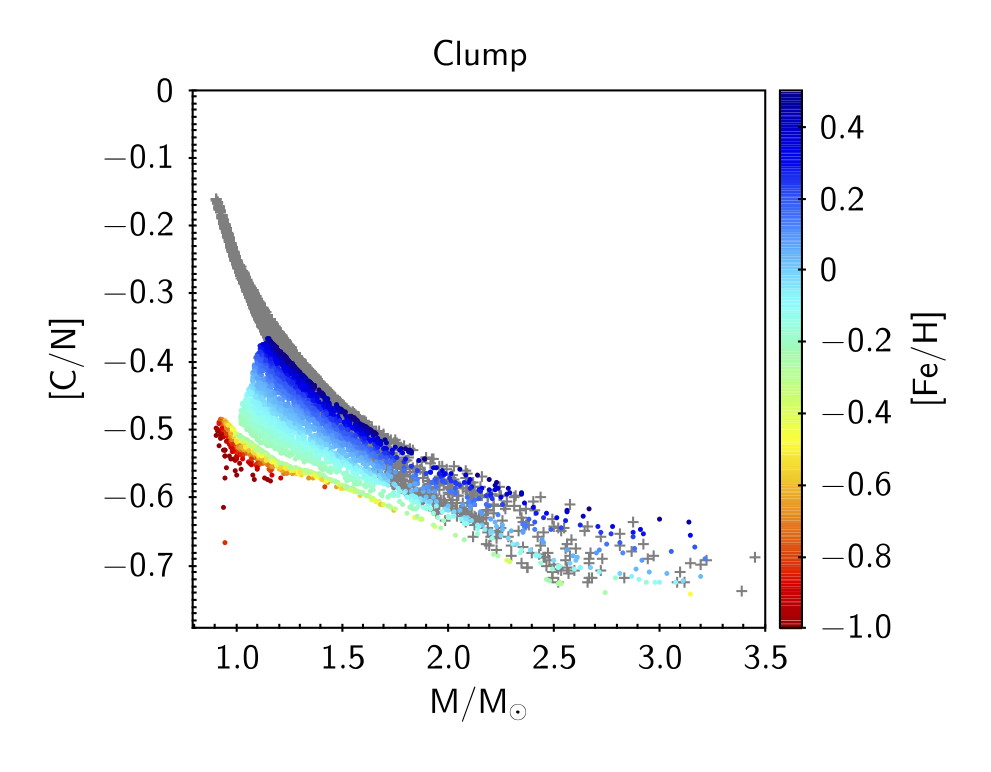}
         \caption{Surface abundance of [C/N] as a function of stellar ages (left panels) and stellar masses (right panels), colour-coded by metallicity, for a synthetic thin disc computed with the BGM. Stars have been selected to be in the clump according to their $\Delta\Pi_{\ell=1}$, as well as before and after the RGB-bump according to their log(g) values.  Simulations including or not the effect of thermohaline instability are shown as colour-dots and gray crosses respectively.}
         \label{Fig_popClumpage}
   \end{figure*}

Figure \ref{Fig_popClump} shows the surface chemical properties of thin disc stars as a function of stellar mass, including or not the effects of thermohaline mixing. This figure focuses on clump stars selected by their asymptotic period spacing of g-modes $\Delta\Pi_{(\ell=1)}$. As shown by \citet{ChaLag10} and recalled in Part 2, thermohaline mixing changes the surface chemical properties of stars more evolved than the RGB-bump. The efficiency of this mechanism with the metallicity and the stellar mass is also shown in Fig.\ref{Fig_popClump}. Due to the strong effect of thermohaline mixing on its abundances, Helium-3 and $^{12}$C/$^{13}$C are the best indicators to constrain this mechanism during the RGB. Fig.\ref{Fig_popClump} shows also an impact on [C/Fe] and [N/Fe] but less important especially for upper metallicities.\\ 

Very recently, \citet{Masseron16} used stellar models (at given mass and metallicity) to compare directly with the APOGEE observations. They claim that stellar evolution models including transport processes (e.g. thermohaline instability and rotation-induced mixing) overestimate the N-abundance in red-clump stars. However this study only considers models with a given mass and does not account for the range of masses and metallicities that are present in observational data. To compare large surveys including stars at different masses and metallicities, synthetic population analysis is the most efficient method to validate the models by comparing the data with simulations, accounting for a realistic range of mass, metallicities and for observational biaises. A detailed comparison between our synthetic populations and large surveys will be done in the Part II of this series (Lagarde et al in prep.).

\section{Determination of age and mass using [C/N] ratio}
   
Recent studies \citep[e.g.][]{Martig15, Masseron15} have proposed to use [C/N] to determine stellar ages and masses of red-giant stars. However, these studies do not take into account the effects of mixing occurring in the stellar interiors, stellar input physics and possible changes of these relations at different evolutionary stages. Figure \ref{Fig_popClumpage} shows [C/N]-dependency with stellar masses and ages, with and without thermohaline instability. Giant stars are divided into three groups: (1) Lower-RGB stars, stars ascending the red-giant branch before the RGB-bump (log(g)>2.2). These stars do not yet undergo thermohaline mixing ; (2) Upper-RGB : brighter RGB stars with log(g)$\leq$2.2 ; (3) Clump stars selected according to their $\Delta\Pi_{(\ell=1)}$ values. 

Considering standard stellar evolution models only (gray dots on Fig. \ref{Fig_popClumpage}), [C/N] seems to be a good proxy to determine stellar masses along the red giant branch and during the He-burning phase. Standard models show a more important dispersion of [C/N] with stellar age than with mass for the first ascent red-giant stars (lower-RGB and upper-RGB). Although [C/N] ratio at the stellar surface is directly related to the stellar properties (mass or metallicity), we point out the difficulty to determine an accurate stellar age from these properties. As discussed by \citet{Lebreton14a} stellar evolution models are still affected by several uncertainties (e.g. [Fe/H], $\alpha$-enhancement, solar mixture, initial He-abundances, transport processes) which can significantly affect the determination of stellar ages from the chemical properties of stars. 

    \begin{figure}[ht!]
   \centering
   \includegraphics[width=0.7\hsize,clip=true,trim= 0.8cm 0.5cm 0.8cm 0.4cm]{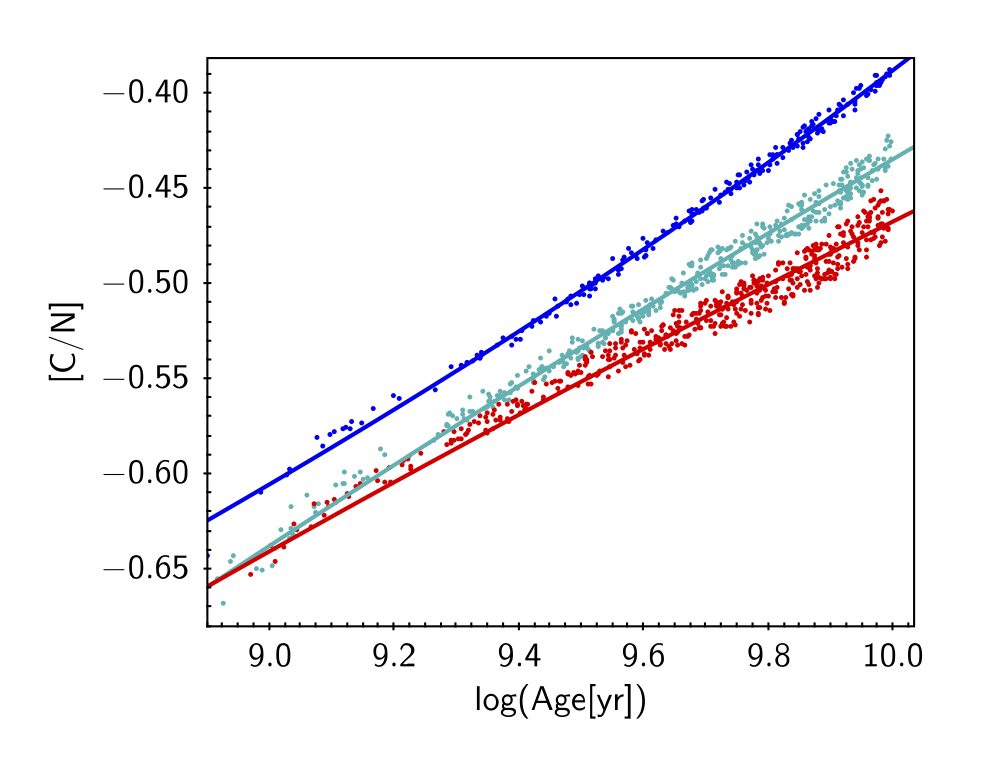}
   \includegraphics[width=0.7\hsize,clip=true,trim= 0.8cm 0.5cm 0.8cm 0.4cm]{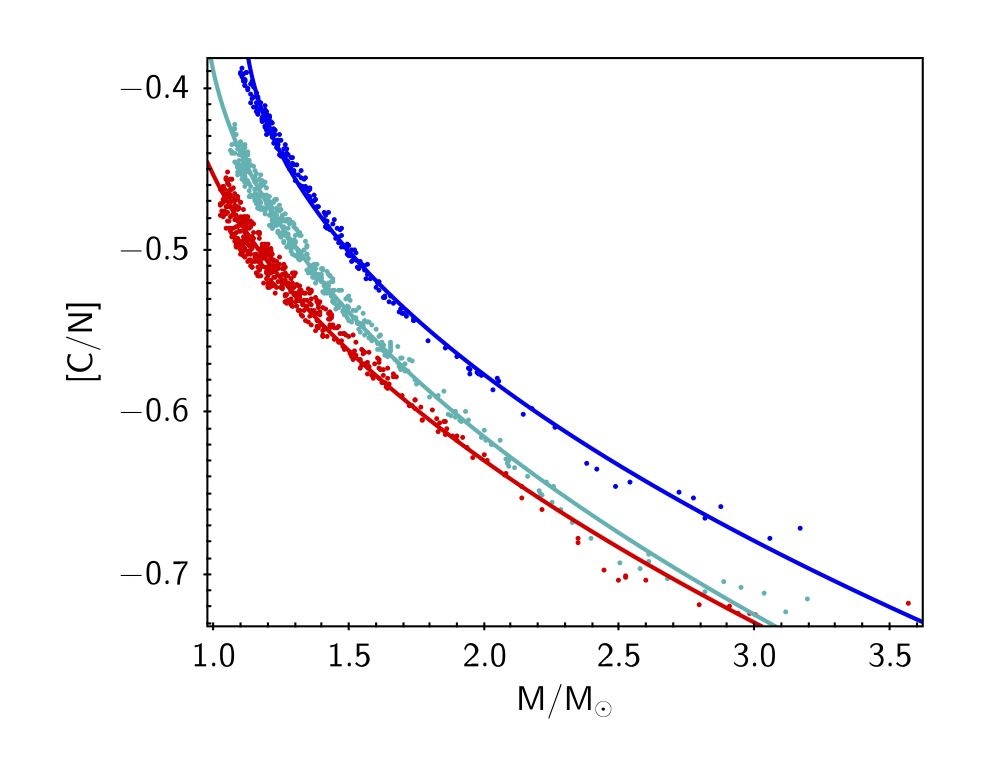}
   
         \caption{Surface abundance of [C/N] for clump stars as a function of stellar ages (top panel) and stellar masses (bottom panel) for a synthetic thin disc computed with the BGM. Stars are divided in three metallicity-bins: -0.20$\leq$ [Fe/H] $\leq$ -0.10 (blue dots), -0.05 $\leq$ [Fe/H] $\leq$ +0.05 (green dots), and +0.15 $\leq$ [Fe/H] $\leq$ +0.25 (red dots). Relations between [C/N] and mass and age (Eq. \ref{rel_mass} and \ref{rel_age}) are also shown with the solid lines.}
         \label{Fig_fitclump}
   \end{figure}
   
In this context, Figure \ref{Fig_popClumpage} (colour-dots) shows the impact of thermohaline instability on the [C/N] vs age and vs  mass diagrams. As thermohaline instability changes the surface abundances in carbon and nitrogen after the bump, large dispersions of [C/N] with mass and age are noticeable for upper-RGB stars. Since this mechanism does not significatively affect the surface chemical properties of clump stars, relationships between [C/N] with masses and ages can be establish. These relationships can be determined at a given metallicity for clump stars contrary to upper-RGB stars.  

Figure \ref{Fig_fitclump} shows thin-disc clump stars divided in three [Fe/H]-bins: (1) lower metallicity (-0.20$\leq$ [Fe/H] $\leq$ -0.10) ; (2) solar metallicity (-0.05 $\leq$ [Fe/H] $\leq$ +0.05) ; (3) higher metallicity (+0.15 $\leq$ [Fe/H] $\leq$ +0.25). In Eq.~\ref{rel_mass} and \ref{rel_age}, we present relations allowing to estimate stellar masses and ages (resp.) of clump stars from [C/N] abundance ratio, in each metallicity range, assuming the stellar physics described above taking into account the effects of thermohaline mixing.

\begin{equation}
\label{rel_mass}
M/M_{\odot} = \left \{
  \begin{aligned}
15.66\cdot[C/N]^{2}+11.27\cdot[C/N]+2.887 \\
  \qquad \text{for} -0.20 \leq [Fe/H] \leq -0.10\\
13.94\cdot[C/N]^{2}+9.554\cdot[C/N]+2.603\\
  \qquad \text{for} -0.05 \leq [Fe/H] \leq +0.05\\ 
17.71\cdot[C/N]^{2}+12.49\cdot[C/N]+3.313\\
  \qquad  \text{for} +0.15 \leq [Fe/H] \leq +0.25
\end{aligned}
\right.
\end{equation}

\begin{equation}
\label{rel_age}
log(Age[yr]) = \left \{
  \begin{aligned}
1.932\cdot[C/N]^{2}+7.904\cdot[C/N]+13.27 \\
  \qquad \text{for} -0.20 \leq [Fe/H] \leq -0.10\\
1.317\cdot[C/N]^{2}+6.329\cdot[C/N]+12.50 \\
  \qquad \text{for} -0.05 \leq [Fe/H] \leq +0.05\\ 
-2.766\cdot[C/N]^{2}+1.845\cdot[C/N]+11.13 \\
  \qquad  \text{for} +0.15 \leq [Fe/H] \leq +0.25
\end{aligned}
\right.
\end{equation}

Importantly, the relation with age depends on the stellar model used. It is then crucial to validate the models and the mixing processes before using ages for Galactic archeology. In a forthcoming paper, we shall discuss the effects of rotation on these relations and estimate the accuracy of age and mass determinations using surface chemical properties.

\section{Conclusions}

In this paper, we have presented a new version of the Besan\c con stellar population synthesis model of the Galaxy including a new grid of stellar evolution models computed with the code STAREVOL. This new stellar grid of single-star evolution models is computed for five metallicities in the mass range between 0.6 and 6.0 M$_\odot$, including the effects of thermohaline instability during the red-giant branch. These models provide the global (e.g. surface gravity, effective temperature,...), chemical (surface properties for 54 stable and unstable species) as well as seismic properties ($\Delta\nu$,$\nu_{max}$,$\Delta\Pi_{(\ell=1)}$). 
Thermohaline mixing occuring in  thin disc giants has been shown to produce measurable effects on the chemical properties, in particular on $^{12}$C/$^{13}$C and [C/N] ratios. Stellar evolution models at different $\alpha$-enhancement are being computed to study the older populations, such as the thick disc, the halo and the bulge, and will be presented in a forthcoming paper.

By comparing the BGM simulations with observations from large spectroscopic and seismic surveys, we are able to constrain the physics of the transport processes occurring in stellar interiors.
Red giants observed from asteroseismology can be now used as new cosmic clock, allowing  age calibration from chemical observations. Applying the new version of the BGM, we derive mean relations between [C/N] and age, usable to estimate ages for thin-disc red-clump giants, knowing their metallicity. 
Contrarily to previously derived relationships, ours take into account the natural spread in mass and metallicity of the underlying population, and allows to include selection biases in the surveys. In a forthcoming paper of this series, we shall investigate the impact of different prescriptions for thermohaline instability on these relations, as well as the effect of rotation-induced mixing.  

Thanks to WEAVE, 4MOST, and PLATO the future looks extremely promising in terms of collecting spectroscopic and seismic data for a large number of stars. The Besan\c con Galaxy model will be a key tool to prepare these future instruments and missions as well as to exploit a large amount of data from Gaia, given a better understanding of stellar and Galactic evolution.

\begin{acknowledgements}
 We acknowledge financial support from "Programme National de Physique Stellaire" (PNPS) of CNRS/INSU, France. N.L. acknowledges financial support from the CNES fellowship. 
 \end{acknowledgements}

\bibliographystyle{aa}
\bibliography{../../../Bibliographie/Reference}
\end{document}